\documentclass[aps,prappl,preprint,superscriptaddress]{revtex4-2}

\RequirePackage{lineno}
\usepackage{graphicx}
\begin{document}


\title{Transverse Magnetic Mode Laser in Photonic Crystal Nanobeam Cavity}

\author{Taesu Ryu}
\affiliation{Department of Optical Engineering, Kongju National University, Cheonan 31080, Republic of Korea}
\author{Hwi-Min Kim}
\affiliation{Department of Physics, Korea Advanced Institute of Science and Technology (KAIST), Daejeon 34141, Republic of Korea}
\author{Sang-Woo Ki}
\affiliation{Department of Optical Engineering, Kongju National University, Cheonan 31080, Republic of Korea}
\author{Yong-Hee Lee}
\affiliation{Department of Physics, Korea Advanced Institute of Science and Technology (KAIST), Daejeon 34141, Republic of Korea}
\author{Jin-Kyu Yang}
\email[]{jinkyuyang@kongju.ac.kr}
\affiliation{Department of Optical Engineering, Kongju National University, Cheonan 31080, Republic of Korea}
\affiliation{Institute of Application and Fusion for Light, Kongju National University, Cheonan, 31080, Republic of Korea}

\date{\today}

\begin{abstract}
We first experimentally demonstrated a transverse magnetic (TM) mode laser in a Photonic crystal (PhC) slab structure at room temperature. This study proposes a PhC nanobeam (NB) cavity to support a high-quality-factor (Q-factor) TM mode. For a large and complete photonic bandgap, the PhC NB structures consist of large air holes in a thick dielectric slab. The PhC NB cavity was optimized numerically for a high-Q-factor TM mode of over 1,000,000 by reducing the radii of the air holes quadratically from the center to the edge of the PhC NB. A single TM mode lasing action was observed in an InGaAsP quantum-well (QW)-embedded optimized PhC NB cavity structure at room temperature via optical pulse pumping, where the QW layer was lightly etched. We believe that the TM mode lasers in PhC NB cavities with a lightly etched QW can be good candidates for a surface plasmon excitation source or a highly sensitive optical sensor.
\end{abstract}


\maketitle

\section{Introduction}
Photonic crystals (PhCs) are employed to control the light–matter interaction using the photonic bandgap (PBG) effect~\cite{yablo87,john87,noda07}. In particular, two-dimensional (2D) PhC slab structures have been considered a useful platform for photonic integrated circuits owing to the three-dimensional (3D) perfect guiding~\cite{chut99,lee02}, easy fabrication~\cite{souk02,hou18}, and scalability of the structure~\cite{joan97}. However, most of the studies conducted on the topic have been related to the transverse electric (TE) mode because of the existence of large PBG for TE polarization and weak coupling between TE and transverse magnetic (TM) polarizations in a thin slab~\cite{john99}.
Since Notomi’s report, a nanobeam (NB) structure has been substituted for a 2D PhC slab structure owing to not only the ultra-high quality factor (Q-factor) but also the smallest possible dielectric cavity~\cite{notomi08,deotare09,yang15}. There are several advantages of the NB cavity, for example, the simple structure for high Q-factor cavity~\cite{deotare09,yang15}, low laser threshold~\cite{kim11,jeong13}, high-density integration~\cite{deotare13,yang21}, ultra-low power optical switching~\cite{shak14,dong22}, and easy integration with the silicon waveguide geometry~\cite{lee17}. Hence, various sensor applications of NB cavities have been demonstrated, such as refractive index sensing~\cite{wang10,qiao18,yang19}, nanoprobe for biosensing~\cite{sham13}, optomechanical sensing~\cite{wu14}, and magnetic field sensing~\cite{du17}. It was reported that thick NB structures can have high Q-factor resonant modes with both TE and TM polarizations~\cite{mc11}. In our previous report, we proposed that a thick NB cavity with a horizon air gap can be a good candidate for ultra-sensitive refractive index sensing~\cite{yang19}. However, to the best of our knowledge, there has been no experimental demonstration of TM mode lasers in NB slab structures because of discouragement of the coupling to the TM mode in the compressive strain at the quantum wells (QWs)~\cite{Hwang00}. There are properties peculiar to TM polarization, such as surface plasmon excitation and strong confinement in the horizontal air gap. In this study, we propose an NB cavity with a thick slab that consists of a 1D array of air holes with quadratic size modification. The numerical results show that an NB structure with a thick slab has a large and complete PBG. In addition, we first experimentally demonstrated a single TM mode laser operation in an optimized NB cavity with an InGaAsP multiple QW layer which was lightly etched for sensing applications. The lasing mode was confirmed by numerical simulation based on a scanning electron microscopy (SEM) image of the sample. We believe that the TM-mode NB laser is a good candidate for a compact on-chip TM-polarization light source for surface plasmon excitation and has various sensor applications. 

\section{Design and Optimization}
\begin{figure}[t]
\includegraphics[width=\textwidth]{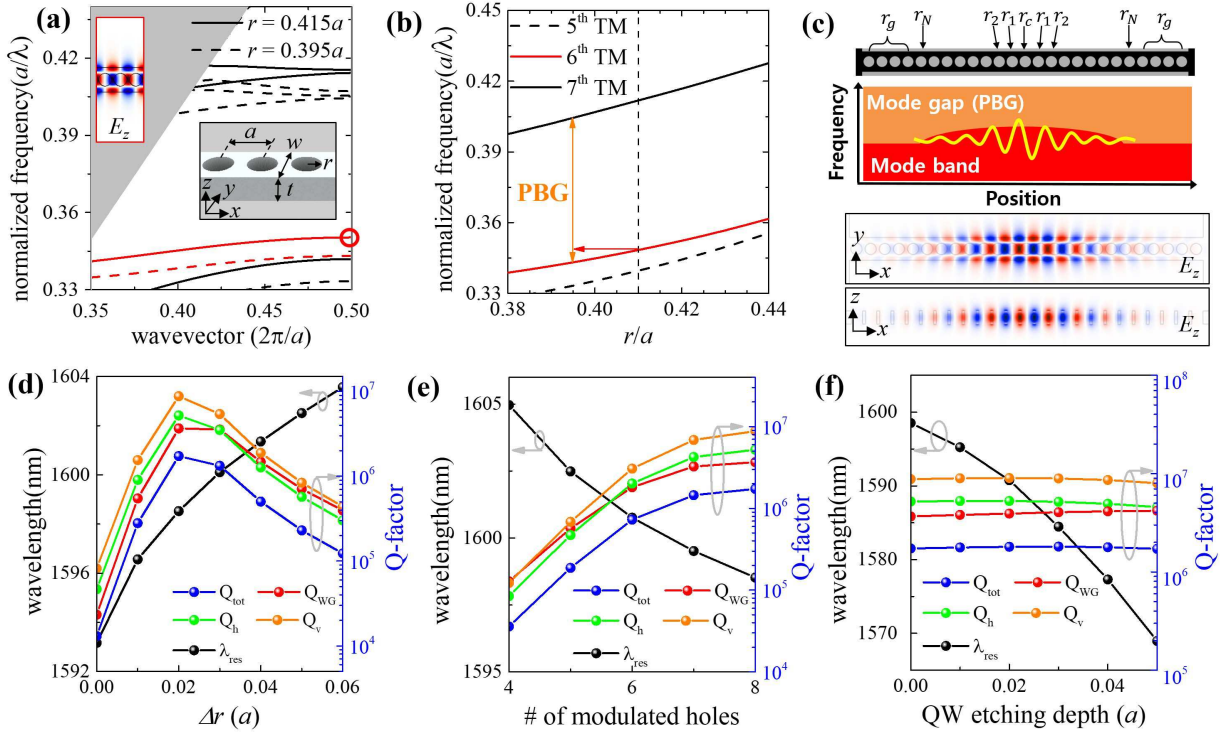}
\caption{Optimization of NB cavity. (a) Photonic band structure of NB with $r/a$ = 0.415 (solid lines) and $r/a$ = 0.395 (dashed lines). (b) Normalized frequency of the TM BE mode ($k = \pi/a$) as a function of the radius of the air hole. (c) $E_z$ field profile of TM cavity mode at the horizontal plane (upper) and vertical plane (lower). Wavelength and Q-factor of TM cavity mode as a function of (d) maximum difference of radii, (e) number of modulated holes, and (f) QW etching depth. The left inset in (a) indicates the $E_z$ field profile of the 6$^{\text{th}}$ TM BE mode marked with the red circle, and the right inset in (a) indicates the schematic image of the NB structure. The upper insets in (c) show the cut-view of the NB cavity and the strategy of light confinement in the cavity by reducing air holes quadratically. \label{fig1}}
\end{figure}

The photonic band structures of NB were calculated using the plane wave expansion (PWE) method~\cite{joan08}. The structural parameters are indicated in the right inset of Figure~\ref{fig1}(a). The period ($a$), thickness ($t$), width ($w$), and refractive index of the slab were set as 563 nm, 520 nm, 1.6$a$, and 3.25, respectively.  Figure~\ref{fig1}(a) shows the band structures for two different hole sizes ($r/a$ = 0.415 and 0.395). For each case, a wide PBG was observed between the 6$^{\text{th}}$ TM band (red solid line) and the 7$^{\text{th}}$ TM band (black solid line). The normal electric field ($E_z$) profile of the 6$^{\text{th}}$ TM band-edge (BE) mode at the wavevector, $k$ = $\pi/a$ (red circle), is shown in the inset of  Figure~\ref{fig1}(a). Subsequently, we obtained the frequency of the TM BE mode as a function of the radius. As shown in  Figure~\ref{fig1}(b), as the radius increases, the TM bands blue-shift owing to the reduction in dielectric materials. Therefore, if an NB with a large air hole is surrounded by an NB with small air holes, the 6$^{\text{th}}$ TM BE mode can be localized in the region of large air holes owing to the PBG effect of the NB with small air holes. Moreover, the 6$^{\text{th}}$ TM BE mode lies inside the TE pseudo-bandgap (Figure S1, Supplementary Material); therefore, the 6$^{\text{th}}$ TM BE mode can be strongly confined by a complete PBG effect.
The TM cavity mode of the NBs is optimized using the 3D finite-difference time-domain (3D FDTD) method. The NB cavity in  Figure~\ref{fig1}(c) is composed of 25 air holes, which are gradually tapered according to the following equation~\cite{yang19},

\begin{equation}
r_i = r_c -  \it{\Delta r} \times i^{\text{2}} / \left( N+1 \right) ^{\text{2}}
\end{equation}
where $r_i$ is the radius of the $i^{\text{th}}$ air hole, $r_c$ is the radius of the center air hole, $\it{\Delta r}$ is the maximum difference between the air hole radii in the NB, $N$ is the number of modulated air holes on one side, and $i$ is the index of the air hole position from the center. To further reduce the propagation loss along the waveguide, four air holes were added at the end of the NB, where the radius $r_g$ was the same as $r_N$.  Figure~\ref{fig1}(c) shows $xy-$ and $xz-$cut views of the $E_z$ field profile of the optimized TM cavity mode. The amplitude of the $E_z$ field has a maximum at the center and gradually decreases along NB.

First, we investigated the effect of $\it{\Delta r}$ on the Q-factor (Q$_{\text{tot}}$) at fixed $N$. The total cavity loss (1/Q$_{\text{tot}}$) is decomposed into waveguide propagation loss (1/Q$_{\text{WG}}$), vertical scattering loss (1/Q$_{\text{v}}$), and horizontal scattering loss (1/Q$_{\text{h}}$). For $N$ = 8, the Q-factor was maximized at approximately 2,000,000 at $\it{\Delta r}$ = 0.02$a$, owing to the minimum optical loss in all channels, as shown in  Figure~\ref{fig1}(d). Subsequently, we set $\it{\Delta r}$ to 0.02$a$ and calculated the Q-factor as a function of $N$, as shown in  Figure~\ref{fig1}(e). The Q-factor increased and saturated as $N$ increased, owing to the saturation of the cavity loss in all channels. We also investigated the Q-factor and resonant wavelength as functions of the QW etching depth. The Q-factor remained unchanged but the wavelength shifted significantly. This implies that the electric field intensity is strongly localized in the QW etching region, which is advantageous for highly sensitive optical sensing~\cite{lee10a,lu12}.

\section{Fabrication}
\begin{figure}[tp]
\includegraphics[width=0.8\textwidth]{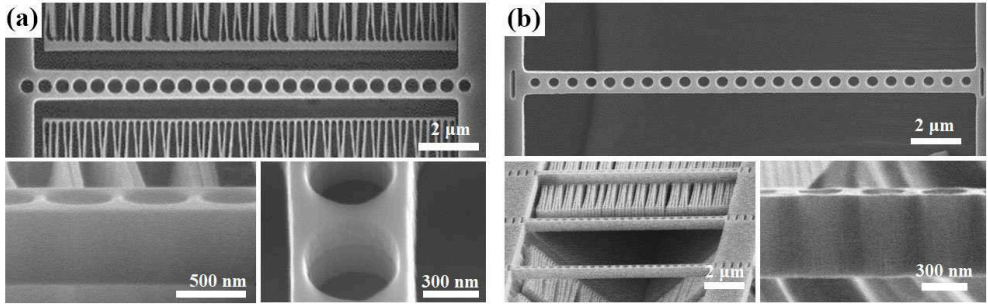}
\caption{SEM images of the fabricated InGaAsP QW-embedded NB cavity samples. (a) InP-NB slab sample. (b) InGaAsP-NB slab sample. 
\label{fig2}}
\end{figure}

The optimized NB cavities were prepared using a standard fabrication technique for III-V semiconductor nanolasers~\cite{park04,moon16}. In this study, we fabricated NB cavities on two types of InGaAsP-QW wafers. One was the InP/InGaAsP-QW slab on the InGaAs sacrificial layer, and the other was an InGaAsP/InGaAsP-QW slab on the InP sacrificial layer. We named the NB cavities fabricated on the InP/InGaAsP-QW wafer and InGaAsP/InGaAsP-QW wafers InP-NBs and InGaAsP-NBs, respectively. The fabrication process was identical for both wafers, except for the final wet-etching step (Figure S2, Supplementary Material). First, poly-methyl methacrylate (PMMA) was spin-coated onto the wafer. Next, photonic crystal patterns were defined using electron-beam lithography. The PMMA layer was hardened by electron beam irradiation and Ar ion milling, and the pattern was transferred to wafers by chemically assisted ion-beam etching (CAIBE). The PMMA layer was removed by O$_2$ plasma ashing. Finally, for the InP-NBs, the membrane was subsequently released by two etching processes. The first wet etching process involved removing the InGaAs sacrificial layer underneath the InP slab using a mixture of succinic acid solution (C$_4$H$_6$O$_4$) and hydrogen peroxide (H$_2$O$_2$) in a volume ratio of 20:1 at room temperature. A second wet etching was performed to selectively remove the InGaAsP QW layer with a mixture of citric acid solution (C$_6$H$_8$O$_7$) and hydrogen peroxide in a volume ratio of 5:1 at a low temperature of approximately 4$^{\circ}$C~\cite{jang15,pram17}. From the SEM images in Figure~\ref{fig2}(a), the thickness of the InP was estimated to be approximately 520 nm, and the QW etching depth was approximately 17 nm. It is worth noting that the QW layer is atomic flat owing to the epitaxy growth, so the thickness of the etched QW is also ultra-uniform, which can enhance the sensitivity of the refractive index of the surrounding media by monitoring the frequency of the TM mode~\cite{yang19,lee10a,jang15}. For the InGaAsP-NBs, a single wet etching process was performed with 20\% diluted HCl solution to remove the InP sacrificial layer. The thickness of the InGaAsP slab was approximately 530 nm, as shown in Figure~\ref{fig2}(b).

\section{Characterization and Discussion}
We optically characterized the fabricated devices with a 980 nm laser diode at room temperature, where the pulse duration and width were 2 $\mu$s and 10 ns, respectively. The pump beam was focused on the device via an objective lens at 40$\times$ magnification. The pump spot size was estimated as 3 $\mu$m in diameter. Photoluminescence (PL) was collected with the same objective lens and delivered to the monochromator to analyze the spectral characteristics of the lasing mode (Figure S3, Supplementary Material).
\begin{figure}[tp]
\includegraphics[width=\textwidth]{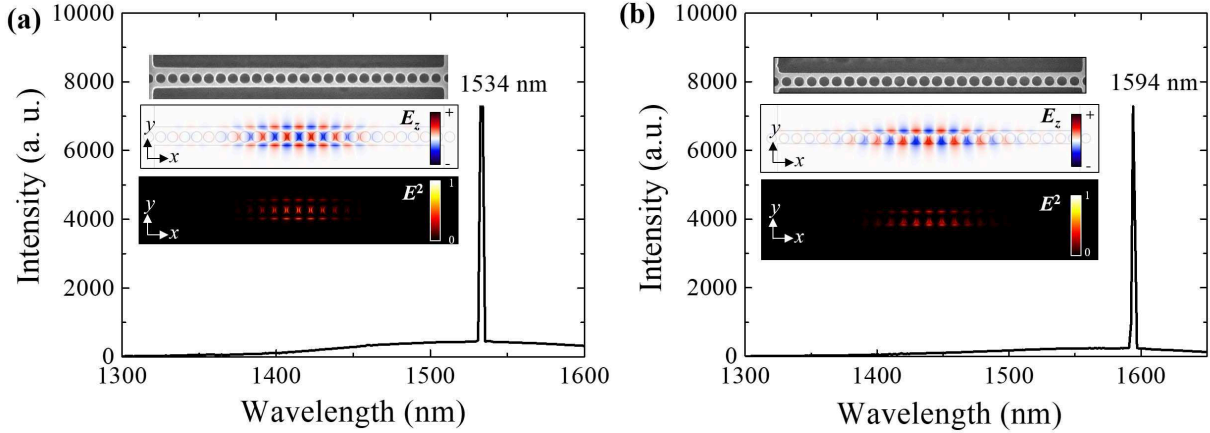}
\caption{Lasing spectra from the InP-NB samples. (a) Lasing spectrum from the TM mode in the symmetric NB sample. (b) Lasing spectrum from the TM mode in the asymmetric NB sample. The inset images are the sample SEM image, the calculated $E_z$ field, and $E^2$ profile of the TM mode from the top to the bottom, respectively. 
\label{fig3}}
\end{figure} 

We observed single-mode operation from both InP-NBs and InGaAsP-NBs. Figure~\ref{fig3} shows the lasing spectra of the two InP-NBs at room temperature. Single-lasing peaks were observed at 1534 and 1594 nm. From the SEM image, the structural parameters were estimated to be: $a$ = 563 nm, $r_c$ = 0.415$a$, $r_g$ = 0.399$a$, and $w$ = 1.60$a$. With 3D FDTD, we calculated the resonant modes using the estimated parameters to verify the lasing modes. One TM mode with a high Q-factor was found near a lasing wavelength of 1534 nm. The calculated wavelength and Q-factor were 1555 nm and 160,000, respectively. In addition, the $E_z$ field profile was identical to that of the target TM BE mode, as shown in the inset of Figure~\ref{fig3}(a). In particular, the electric field intensity was strongly localized at the boundary of the NB, where the QW was lightly etched laterally~\cite{yang19,lee10b}. There was also a TE cavity mode (See Figure S4(a) in the Supplementary Material). However, the calculated wavelength was 1508 nm, which is far from the lasing peak, and the Q-factor was 1,000, which is less favorable for lasing action. We performed the same analysis on the asymmetrical InP-NB laser, as shown in the top inset of Figure~\ref{fig3}(b). The estimated structural parameters were as follows: $a$ = 571 nm, $r_c$ = 0.407$a$, $r_g$ = 0.349$a$, and $w$ = 1.10$a$. We found a TM mode with a wavelength of 1618 nm and a Q-factor of 17,000, as shown in the middle inset of Figure~\ref{fig3}(b). There was a TE mode at a wavelength of 1441 nm, but the Q-factor was 520 (See Figure S4(b) in the Supplementary Material). Owing to the inherent radiation characteristics of the TM mode, the emission was not captured with an infrared charged-coupled device (IR CCD). In addition, because of the poor thermal conductivity of the QW etching NB cavity, we could not obtain the light-in-light-out (L-L) curve and near-field image at the lasing action. 

\begin{figure}[tp]
\includegraphics[width=\textwidth]{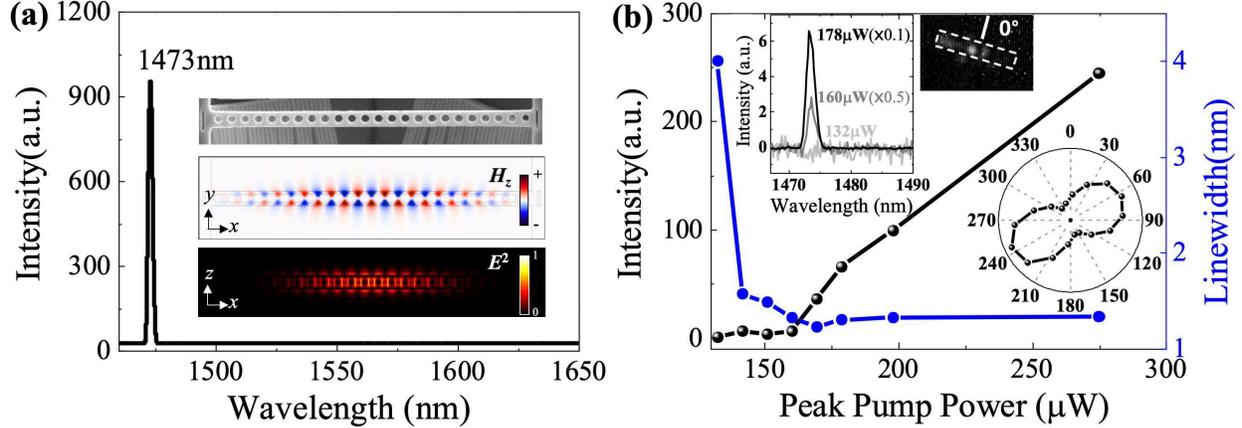}
\caption{Characteristics of lasing action from the InGaAsP-NB sample. (a) Lasing spectrum from the NB sample. (b) L-L curve. The inset images in (a) are the sample SEM image, the calculated $H_z$ field profile of the TE cavity mode in the normal direction, and the $E^2$ profile from the top to the bottom, respectively. The insets in (b) are the spectra at different pumping levels, the near-field image, and the polarization property at lasing action from the left to the right, respectively 
\label{fig4}}
\end{figure}

In addition, we observed the lasing action of the TE mode from InGaAsP-NBs. From the SEM image in the top inset of Figure~\ref{fig4}(a), the structural parameters were estimated as follows: $a$ = 573 nm, $r_c$ = 0.343$a$, $r_g$ = 0.302$a$, and $w$ = 1.12$a$. Here, the thickness of the InGaAsP slab was 530 nm, which is 10 nm thicker than that of the InP slab. With 3D FDTD, a TE mode was found near the lasing wavelength of 1473 nm. The simulated wavelength was 1478 nm, and the Q-factor was approximately 150,000 (See Figure S1(d) in Supplementary Material). There was a TM mode at a wavelength of 1447 nm, but the Q-factor was relatively low at approximately 44,000. In particular, the $H_z$ field profile of the TE mode is well balanced to reduce the optical loss similar to the photonic bound states in the continuum~\cite{gao16}. Figure~\ref{fig4}(b) shows the L–L and linewidth characteristics that have threshold behavior of lasing action. The threshold peak pump power was 160 $\mu$W and the linewidth reduction was observed near the threshold. We characterized multiple InGaAsP-NBs and observed lasing actions primarily from TE cavity modes (See Figure S5 in the Supplementary Material). Multimode lasing originated from the 3$^{\text{rd}}$ TE, and the 6$^{\text{th}}$ TM cavity mode was also observed, which was confirmed by 3D FDTD analysis with estimated parameters (Figure S5(c) in the Supplementary Material and Table S1). 

One of the promising applications of TM lasers is as surface plasmon polaritons (SPPs) excitation source for plasmonic waveguides. We performed a 3D FDTD simulation to excite the SPP with a TM mode NB. The proposed hybrid NB-plasmonic structure consists of three parts: the center part is an NB with $r/a$ = 0.35 as a resonator; the right part is another NB with $r/a$ = 0.3 as a mirror; and the left part is an Au tip-shaped waveguide with a thickness of 0.05$a$ on top of a straight waveguide. Here, the slab thickness ($t$), width ($w$), and lattice constant ($a$) of the NB were set to 530, 696, and 580 nm, respectively. Figure S6 shows the numerical results of the SPP coupling from the TM BE mode. From the side view of the $E^2$ intensity profile, the TM BE mode was successfully coupled to the Au SPP waveguide. 

\section{Conclusion}
We first experimentally demonstrated TM-mode lasers in a thick PhC NB slab structure at room temperature. The NB cavity was numerically optimized by gradually reducing the air holes from the center to the edge of the NB cavity to exist a high-Q TM mode of over 1,000,000. Two types of TM-mode lasers were fabricated on two epitaxial wafers. The first type was InP-NBs, in which the QW was laterally etched to enhance environmental sensitivity, and the second type was InGaAsP-NBs without QW etching. Single-mode lasing action was observed in both InP-NB and InGaAsP-NB at room temperature via optical pulse pumping. The TM laser mode was confirmed by SEM image-based numerical simulations. We believe that TM-mode lasers with a lightly etched QW can be an efficient SPP excitation source and be used as a highly sensitive optical sensor.

\begin{acknowledgments}
This work was supported by the Basic Science Research Program through the National Research Foundation of Korea (NRF) funded by the Ministry of Science and ICT (No. 2020R1A2C1014498) and a research grant from Kongju National University in 2020. 
\end{acknowledgments}



\widetext
\clearpage
\textbf{\large Supplementary Information: Transverse Magnetic Mode Laser in Photonic Crystal Nanobeam Cavity}

\begin{center}
\textbf{Taesu Ryu$^1$, Hwi-Min Kim$^2$, Sang-Woo Ki$^1$, Yong-Hee Lee$^2$, and Jin-Kyu Yang$^{1,3,*}$}

\textit{$^{1}$ Department of Optical Engineering, Kongju National University, Cheonan 31080, Republic of Korea}

\textit{$^{2}$Department of Physics, Korea Advanced Institute of Science and Technology (KAIST), Daejeon 34141, Republic of Korea}

\textit{$^{3}$Institute of Application and Fusion for Light, Kongju National University, Cheonan, 31080, Republic of Korea}
\end{center}

\date{\today}

\setcounter{equation}{0}
\setcounter{figure}{0}
\setcounter{table}{0}
\setcounter{page}{1}
\makeatletter
\renewcommand{\theequation}{S\arabic{equation}}
\renewcommand{\thefigure}{S\arabic{figure}}
\renewcommand{\bibnumfmt}[1]{[S#1]}
\renewcommand{\citenumfont}[1]{S#1}


\begin{figure}[htb]
\includegraphics[width=\textwidth]{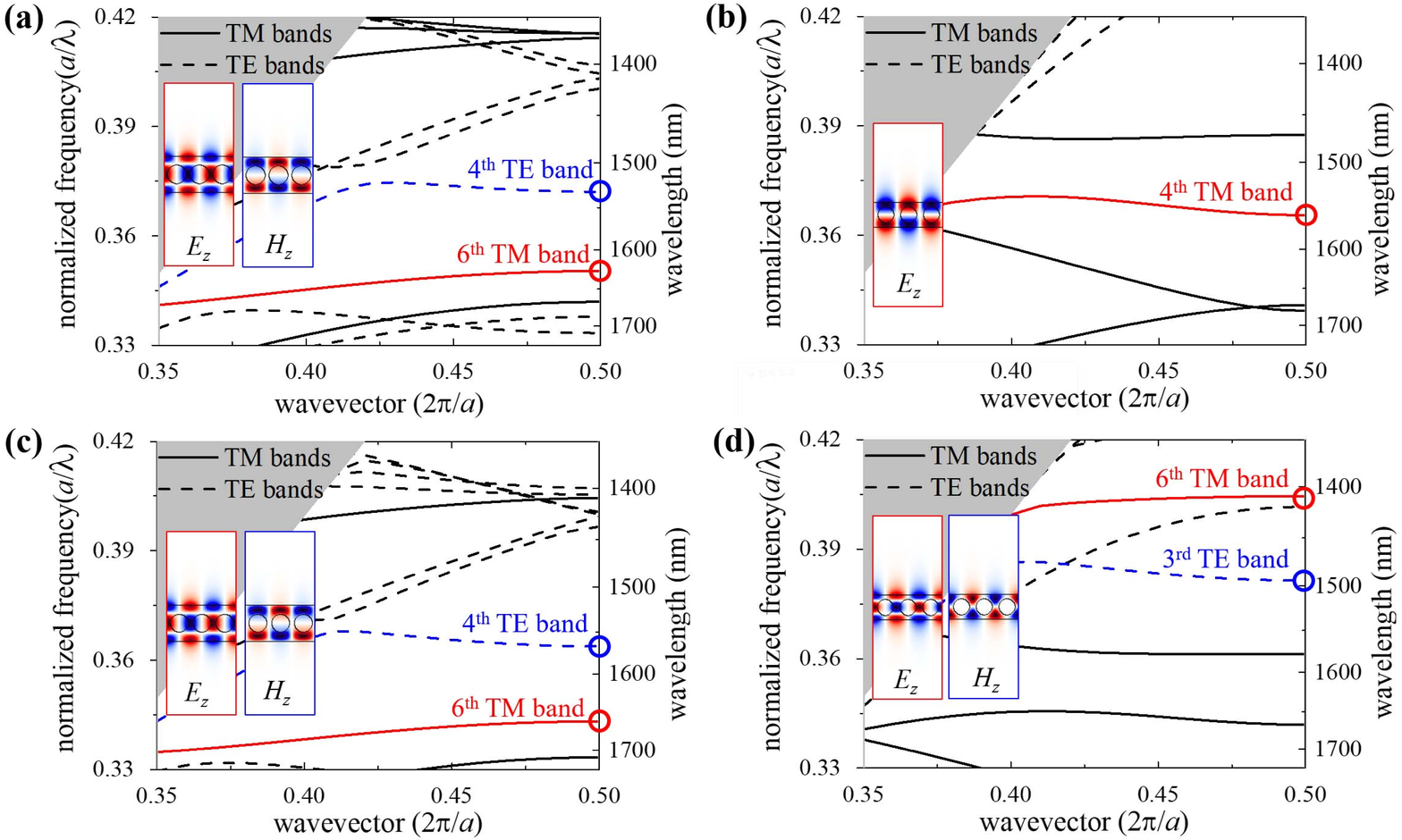}
\caption{Photonic band structures of various NB structures. (a) $r$ = 0.415 $a$, $w$ = 1.60 $a$ (b) $r$ = 0.407 $a$, $w$ = 1.10 $a$ (c) $r$ = 0.395 $a$, $w$ = 1.60 $a$, (d) $r$ = 0.350 $a$, $w$ = 1.10 $a$. The inset images indicate the field profiles of 4$^{\text{th}}$ (or 3$^{\text{rd}}$) TE mode ($H_z$) and 6$^{\text{th}}$ TM mode ($E_z$) marked with red and blue circles, respectively.     
\label{figS1}}
\end{figure}


\begin{figure}[htb]
\includegraphics[width=0.8\textwidth]{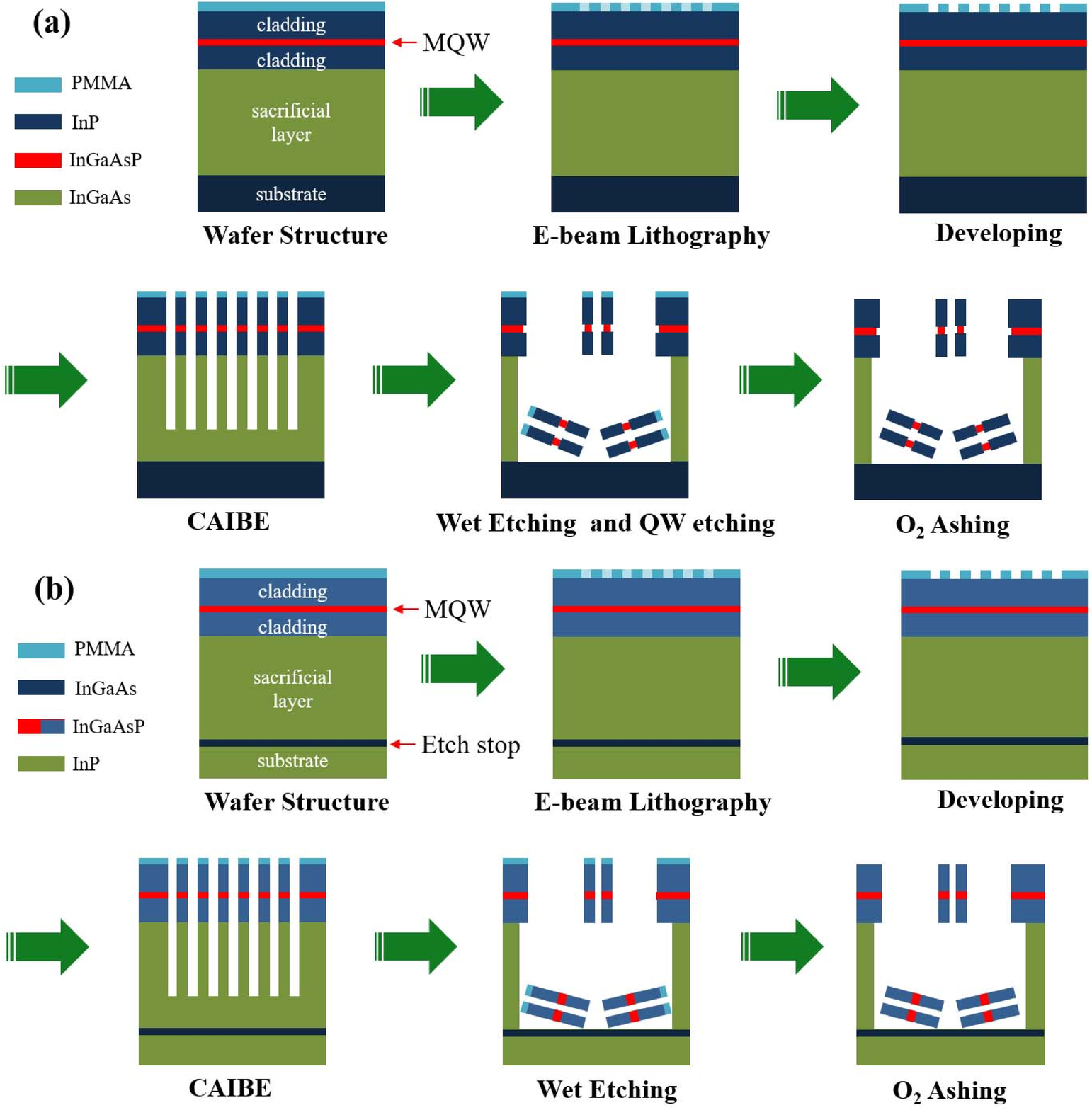}
\caption{Fabrication procedure (a) NB cavity in free-standing InP slab including QW etching process. (b) NB cavity in free-standing InGaAsP slab without QW etching.
\label{figS2}}
\end{figure}


\begin{figure}[htb]
\includegraphics[width=0.8\textwidth]{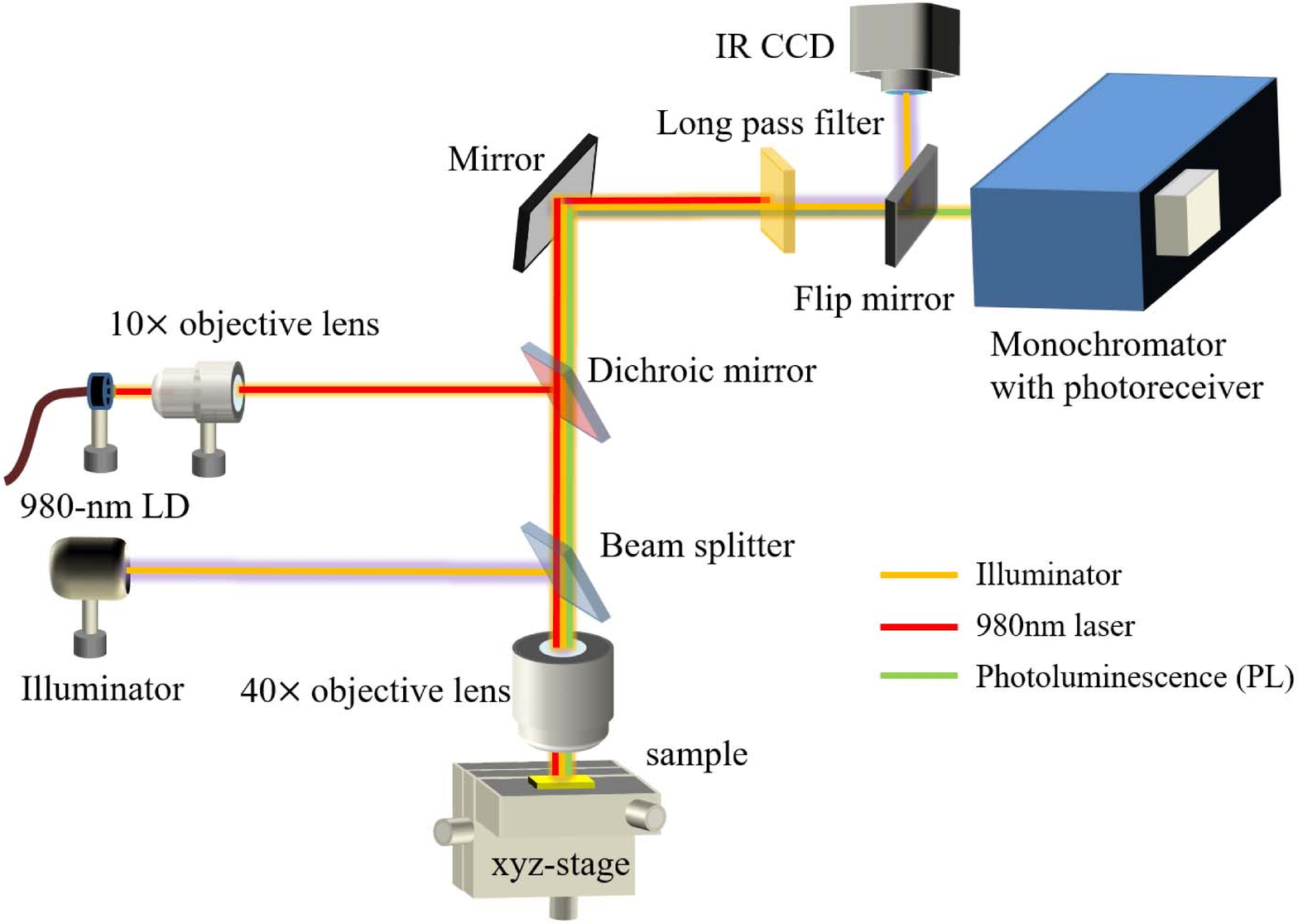}
\caption{Experimental setup.
\label{figS3}}
\end{figure}

\begin{figure}[tb]
\includegraphics[width=\textwidth]{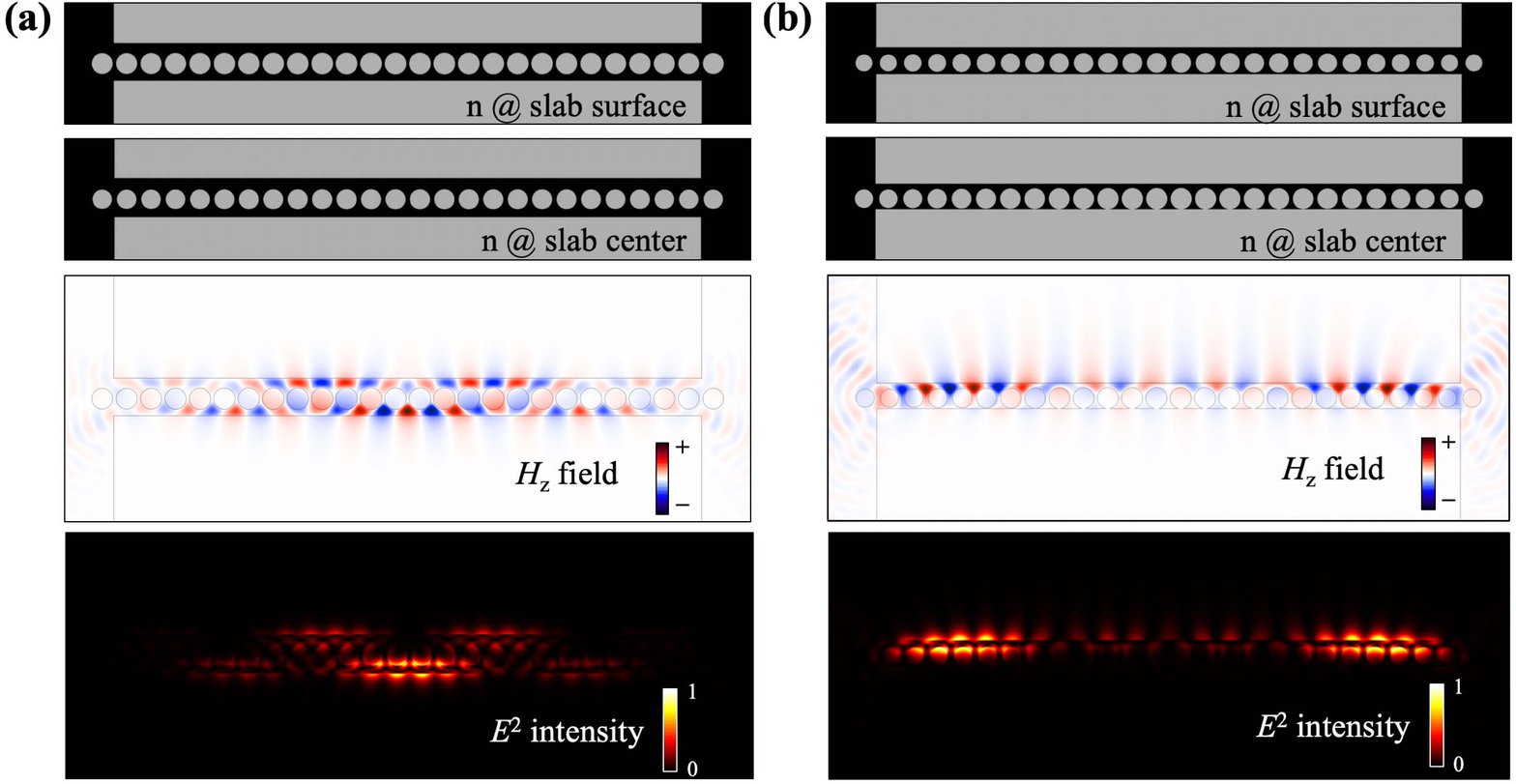}
\caption{
FDTD simulation results of TE resonant modes in the InP-NB cavity using the estimated parameters. (a) symmetric wide NB (b) asymmetric narrow NB. The images indicate the refractive index profiles at the slab surface and at the slab center, $H_z$ field profile of TE resonant mode and $E^2$ profile at the slab center from the top to the bottom.
\label{figS4}}
\end{figure} 

\begin{figure}[htb]
\includegraphics[width=\textwidth]{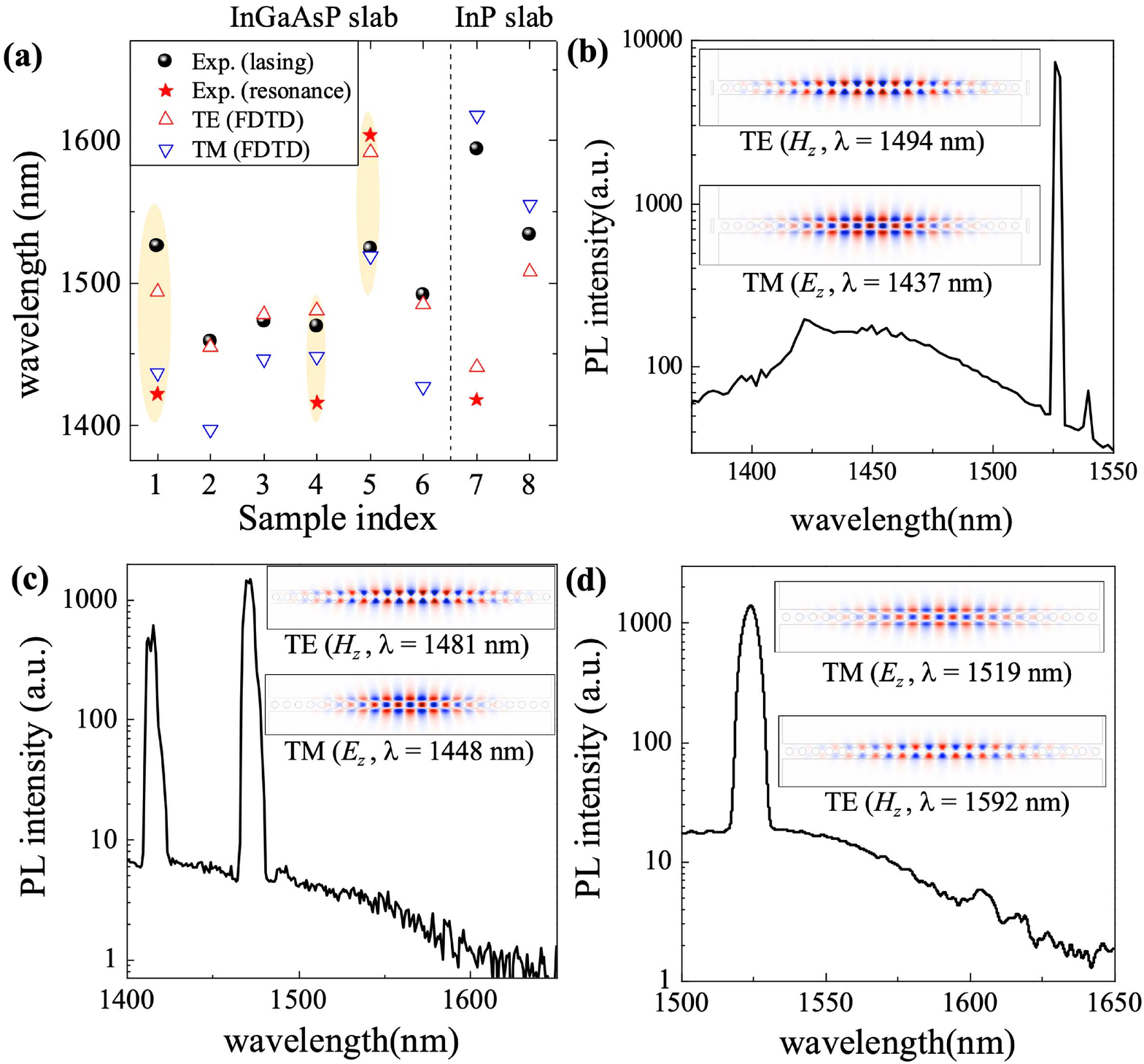}
\caption{
Characteristics of lasing in NB cavity samples. (a) wavelength comparison between the experimental lasing peaks and calculated resonances using the estimated parameters. Lasing spectra at (b) the 1$^{\text{st}}$ sample (c) the 4$^{\text{th}}$sample (d) the 5$^{\text{th}}$ sample. The inset images in (b) $\sim$ (d) shows the calculated field profiles of TE and TM mode at resonant wavelengths. 
\label{figS5}}
\end{figure} 

\newpage
\begin{table}[htb]
\caption{Structural parameters of the samples, lasing peaks, and calculation results using the estimated structural parameters.\label{tb1}}
\begin{ruledtabular}
\begin{tabular}{c|c|c|c|c|c|c|c|c|c}
Sample 	& 	$a$[nm]		&	$w/a$	&	$r_c/a$	&	$r_g/a$	&	$t$[nm]	&	$\lambda_{\text{exp}}$[nm]	&	$\lambda_{\text{FDTD}}$[nm]	&	Q-factor	&	mode	\\
\hline
1&	566&	1.10&	0.321&	0.286&	530&	1526&	1494&	239,000&		3$^{\text{rd}}$ TE	\\
2&	573&	1.04&	0.329&	0.285&	530&	1459&	1455&	87,400&		3$^{\text{rd}}$ TE	\\
3&	573&	1.12&	0.343&	0.302&	530&	1473&	1473&	149,000&		3$^{\text{rd}}$ TE	\\
4&	573&	1.12&	0.342&	0.297&	530&	1470&	1481&	124,000&		3$^{\text{rd}}$ TE	\\
5&	595&	1.12&	0.313&	0.285&	530&	1524&	1519&	435,000&		6$^{\text{th}}$ TM	\\
6&	595&	1.02&	0.329&	0.283&	530&	1492&	1485&	9,600&		3$^{\text{rd}}$ TE	\\
7&	571&	1.10&	0.407&	0.349&	520&	1594&	1618&	17,000&		6$^{\text{th}}$ TM	\\
8&	563&	1.60&	0.415&	0.399&	520&	1534&	1555&	160,000&		6$^{\text{th}}$ TM	\\
\end{tabular}
\end{ruledtabular}
\end{table}

\begin{figure}[htb]
\includegraphics[width=\textwidth]{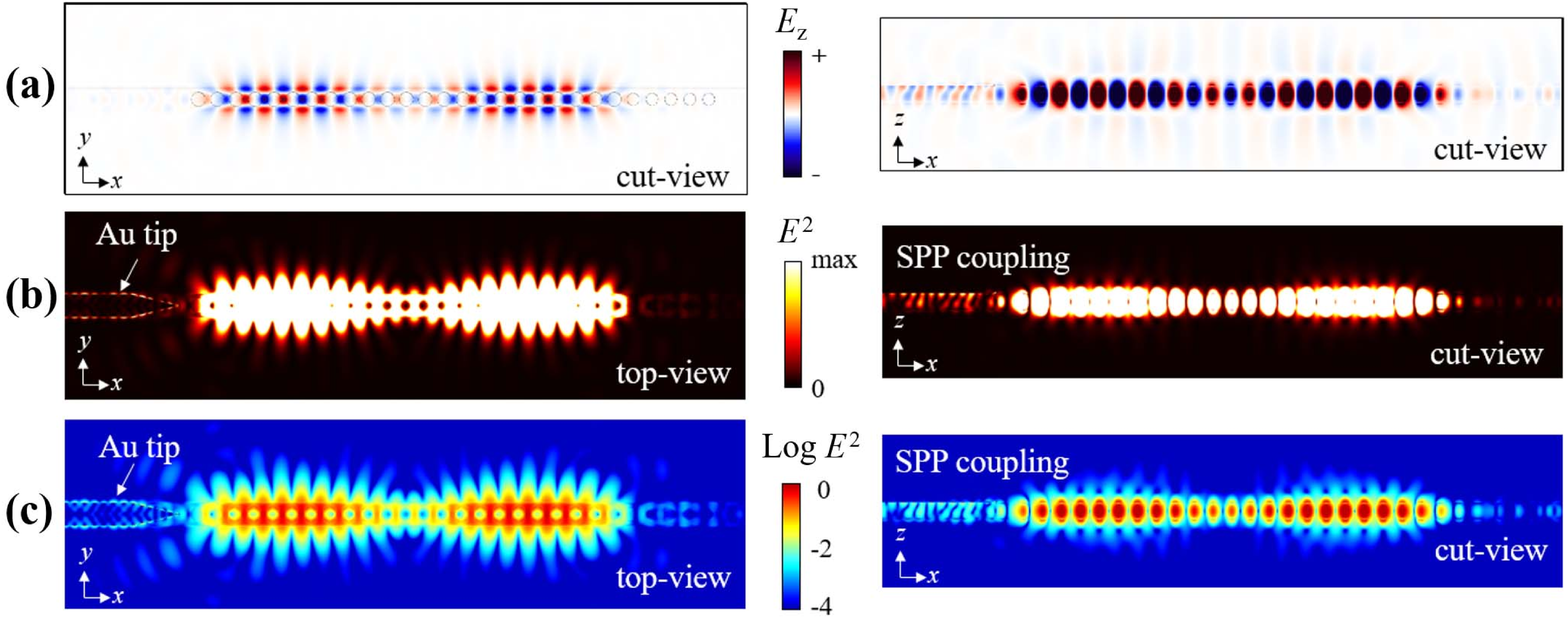}
\caption{
Numerical results of the high-Q TM resonant mode in NB slab with Au strip waveguide. (a) $E_z$ field profile, (b) $E^2$ profile, (c) logarithmic scale of $E^2$ profile. The left and right images show the cut views of the slab center and the normal to the slab, respectively.
\label{figS6}}
\end{figure}


\begin{thebibliography}{99}
\bibitem{yablo87} E. Yablonovitch, \emph{Phys. Rev. Lett.} \textbf{58}, 2059-2062 (1987).
\bibitem{john87} S. John, \emph{Phys. Rev. Lett.} \textbf{58}, 2486-2489 (1987).
\bibitem{noda07} S. Noda, M. Fujita, and T. Asano, \emph{Nature Photon.} \textbf{1}, 449-458 (2007).
\bibitem{chut99} A. Chutinan and S. Noda, \emph{Appl. Phys. Lett.} \textbf{75}, 3739-3741 (1999).
\bibitem{lee02} W. Lee, S. A. Pruzinsky, and P. V. Braun, \emph{Adv. Mater.} \textbf{14}, 271-274 (2002).

\bibitem{souk02} C. M. Soukoulis, \emph{Nanotechnology} \textbf{13}, 420-423 (2002).
\bibitem{hou18} J. Hou, M. Li, and Y. Song, \emph{Angew. Chem. Int. Ed.} \textbf{130}, 2571-2581 (2018).
\bibitem{joan97} J. D. Joannopoulos, P. R. Villeneuve, and S. Fan, \emph{Nature} \textbf{386}, 143-149 (1997).
\bibitem{john99} S. G. Johnson, S. Fan, P. R. Villeneuve, J. D. Joannopoulos, and L. A. Kolodziejski, \emph{Phys. Rev. B} \textbf{60}, 5751-5758 (1999).
\bibitem{notomi08} M. Notomi, E. Kuramochi, and H. Taniyama, \emph{Opt. Express} \textbf{16}, 11095-11102 (2008).

\bibitem{deotare09} P. B. Deotare, M. W. McCutcheon, I. W. Frank, M. Khan, and M. Lon$\breve{\text{c}}$ar, \emph{Appl. Phys. Lett.} \textbf{94}, 121106 (2009).
\bibitem{yang15} D. Yang, P. Zhang, H. Tian, Y. Ji, and Q. Quan, \emph{IEEE Photonics J.} \textbf{7}, 1-8 (2015).
\bibitem{kim11} S. Kim, B.-H. Ahn, J.-Y. Kim, K.-Y. Jeong, K. S. Kim, and Y.-H. Lee, Y. H.  \emph{Opt. Express} \textbf{19}, 24055-24060 (2011).
\bibitem{jeong13} K.-Y. Jeong, Y. Hwang, K. S. Kim, M.-K. Seo, H.-G. Park, and Y.-H. Lee, \emph{Nat. Commun.} \textbf{4}, 2822 (2013).
\bibitem{deotare13}  P. B. Deotare, L. C. Kogos, I. Bulu, and M. Lon$\breve{\text{c}}$ar, \emph{IEEE J. Sel. Top. Quantum Electron.} \textbf{19}, 3600210 (2013).

\bibitem{yang21} D. Yang, X. Liu, X. Li, B. Duan, A. Wang, and Y. Xiao, \emph{J. Semicond} \textbf{42}, 023103 (2021).
\bibitem{shak14} A. Shakoor, K. Nozaki, E. Kuramochi, K. Nishiguchi, A. Shinya, and M. Notomi, \emph{Opt. Express} \textbf{22}, 28923-28634 (2014).
\bibitem{dong22} P. Dong, L. Zhang, D. Dai, Y. Shi, \emph{ACS Photonics} \textbf{9}, 197-202 (2022).
\bibitem{lee17} J. Lee, I. Karnadi, J. T. Kim, Y.-H. Lee, and M.-K. Kim, \emph{ACS Photonics} \textbf{4}, 2117-2123 (2017).
\bibitem{wang10} B. Wang. M. A. D$\ddot{\text{u}}$ndar, R. N$\ddot{\text{o}}$tzel, F. Karouta, S. He, and R. W. van der Heijden, \emph{Appl. Phys. Lett.} \textbf{97}, 151105 (2010).

\bibitem{qiao18} Q. Qiao, J. Xia, C. Lee, and G. Zhou, \emph{Micromachines} \textbf{9}, 541 (2018).
\bibitem{yang19} J.-K. Yang, C.-Y. Kim, and M. Lee, \emph{Appl. Sci.} \textbf{9}, 967 (2019).
\bibitem{sham13} G. Shambat, S. R. Kothapalli, J. Provine, T. Sarmiento, J. Harris, S. S. Gambhir, and J. Vu$\breve{\text{c}}$kovi$\acute{\text{c}}$, \emph{Nano Lett.} \textbf{13}, 4999-5005 (2013).
\bibitem{wu14} M. Wu, A. C. Hryciw, C. Healey, D. P. Lake, H. Jayakumar, M. R. Freeman, J. P. Davis, and P. E. Barclay, \emph{Phys. Rev. X} \textbf{4}, 021052 (2014).
\bibitem{du17} H. Du, G. Zhou, Y. Zhao, G. Chen, and F. S. Chau, \emph{Appl. Phys. Lett.} \textbf{110}, 061110 (2017).

\bibitem{mc11} M. W. McCutcheon, P. B. Deotare, Y. Zhang, and M. Lon$\breve{\text{c}}$ar, \emph{Appl. Phys. Lett.} \textbf{98}, 111117 (2011).
\bibitem{Hwang00} J. K. Hwang, H. Y. Ryu, D. S. Song, I. Y. Han, H. K. Park, D. H. Jang, and Y. H. Lee, \emph{IEEE Photonics Technol. Lett.} \textbf{12}, 1295-1297 (2000).
\bibitem{joan08} J. D. Joannopoulos, S. G. Johnson, J. N. Winn, and R. D. Meade, \textit{Molding the Flow of Light} (Princeton
University Press, Princeton, NJ, 2008).
\bibitem{lee10a} S. Lee, S. C. Eom, J. S. Chang, C. Huh, G. Y. Sung, and J. H. Shin, \emph{Opt. Express} \textbf{18}, 20638-20644 (2011).
\bibitem{lu12} T. W Lu, P. T. Lin, and P. T. Lee, \emph{Opt. Lett.} \textbf{37}, 569-571 (2012).
\bibitem{park04} H.-G. Park, S.-H. Kim, S.-H. Kwon, Y.-G. Ju, J.-K. Yang, J.-H. Baek, S.-B. Kim, and Y.-H. Lee, \emph{Science} \textbf{305}, 1444-1447 (2004).

\bibitem{moon16} S.-K. Moon, K.-Y. Jeong, H. Noh, and J.-K. Yang, \emph{Appl. Phys. Lett.} \textbf{109}, 241106 (2016).
\bibitem{jang15} H. Jang, I. Karnadi, P. Pramudita, J.-H. Song, K. S. Kim, and Y.-H. Lee, \emph{Nat. Commun.} \textbf{6}, 8276 (2015).
\bibitem{pram17} P. Pramudita, H. Jang, I. Karnadi, H.-M. Kim, and Y.-H. Lee, \emph{Opt. Express} \textbf{25}, 6311-6319 (2017).
\bibitem{lee10b} S. Lee, S. C. Eom, J. C. Chang, C. Huh, G. Y. Sung, and J. H. Shin, \emph{Opt. Express} \textbf{18}, 11209-11215 (2010).
\bibitem{gao16} X. Gao, C. W. Hsu, B. Zhen, X. Lin, J. D. Joannopoulos, and M. Solja$\breve{\text{c}}$i$\acute{\text{c}}$, M.; Chen, H. \emph{Sci. Rep.} \textbf{6}, 31908 (2016).


\end{thebibliography}

\begin{thebibliography}{11}

\end{thebibliography}

\end{document}